\documentclass[a4paper,fleqn,usenatbib]{mnras}
\usepackage[T1]{fontenc}
\usepackage{ae,aecompl}
\usepackage{graphicx}
\usepackage{amsmath}
\usepackage{amssymb}
\usepackage{txfonts}

\newcommand{\ltsima}{$\; \buildrel < \over \sim \;$}
\newcommand{\lsim}{\lower.5ex\hbox{\ltsima}}
\newcommand{\gtsima}{$\; \buildrel > \over \sim \;$}
\newcommand{\gsim}{\lower.5ex\hbox{\gtsima}}

\title[The role of cosmology in modern physics]
{The role of cosmology in modern physics}
\author[B.M. Sch{\"a}fer]
{Bj{\"o}rn Malte Sch{\"a}fer\thanks{e-mail: bjoern.malte.schaefer@uni-heidelberg.de}\\
Astronomisches Recheninstitut, Zentrum f{\"u}r Astronomie der Universit{\"a}t Heidelberg, Philosophenweg 12, 69120 Heidelberg, Germany}

\begin{document}
\pagerange{\pageref{firstpage}--\pageref{lastpage}}
\pubyear{2017}
\maketitle
\label{firstpage}

\begin{abstract}
Subject of this article is the relationship between modern cosmology and fundamental physics, in particular general relativity as a theory of gravity on one side, together with its unique application in cosmology, and the formation of structures and their statistics on the other. It summarises arguments for the formulation for a metric theory of gravity and the uniqueness of the construction of general relativity. It discusses symmetry arguments in the construction of Friedmann-Lema{\^i}tre cosmologies as well as assumptions in relation to the presence of dark matter, when adopting general relativity as the gravitational theory. A large section is dedicated to $\Lambda$CDM as the standard model for structure formation and the arguments that led to its construction, and to the of role statistics and to the problem of scientific inference in cosmology as an empirical science. The article concludes with an outlook on current and future developments in cosmology.
\end{abstract}


\section{Introduction}
Modern cosmology aims to describe the expansion dynamics of the Universe on large scales and the formation and evolution of cosmic structures with physical models. In this endeavour cosmology links fundamental laws of physics, most importantly general relativity as the theory of gravity, with (fluid) mechanics for the formation of structures and statistics to provide a framework for interpreting cosmological observations. 

Experimental results from the last decades have established the $\Lambda$CDM as the standard model for cosmology and have measured the set of cosmological parameters at percent precision. Measurements of the distance redshift relation of cosmological objects from Cepheids on small distances \citep{freedman_hubble_2010, freedman_final_2001} to supernovae on large distances \cite{Perlmutter1998, Riess1998} have established a value of the Hubble constant and the presence of accelerated expansion on large scales \citep[for a review, see][]{weinberg_observational_2013}, while at the same time observations of the cosmic microwave background \citep{1990ApJ...354L..37M, 1991AdSpR..11..181M, 1991AdSpR..11..193S} constrained the spatial curvature to be very small and allowed a precision determination of cosmological parameters, in particular those related to inflation from the statistics of cosmic structures as a function of scale. Even more importantly, the adiabaticity of perturbations ruled out structure generation models based on cosmic defects. Almost in passing the Friedmann-Lema{\^i}tre-Robertson-Walker-class of cosmologies \citep{friedmann_uber_1924, lemaitre_expansion_1931, robertson_kinematics_1935} was able to provide a framework of thermal evolution with the epochs of element formation and atom formation in the early universe. Converging on the current value of the Hubble constant and adopting a nonzero value of $\Lambda$ \citep{carroll_cosmological_1992, carroll_cosmological_2001} it was possible to accommodate old astrophysical objects such as white dwarfs and slowly decaying atomic nuclei within the age of the Universe suggested by the cosmological model.

It is important to realise, however, that cosmology differs substantially from other branches of physics: This concerns in particular the passive role of observers, the process of inference from data and the importance of assumptions on the cosmological model which are difficult to test or which can not be tested at all. Furthermore, gravity as the fundamental interaction in cosmology is peculiar because its formulation is unique if certain physical concepts are accepted, so any observed deviation from the prediction of relativity will not shed doubt on a particular formulation of the gravitational model but directly on these fundamental physical concepts. Given these perspectives, cosmology brings together deductive reasoning (concerning the construction of gravitational theories) with statistical inference for selecting the true model, meaning the most economic model compatible with the data.

In this article I will discuss the relation between cosmology and fundamental physics on one side and the process of statistical inference on the other. Specifically, I intend to discuss the concepts of general relativity in Sect.~\ref{sect_gravity}, their application to cosmology in Sect.~\ref{sect_cosmology} and the construction of $\Lambda$CDM as the standard model of cosmology in Sect.~\ref{sect_lambdacdm}. The peculiar role of statistics in cosmology and the process of statistical inference are treated in Sects.~\ref{sect_statistics} and~\ref{sect_inference}. Cosmic structure formation with its relations to relativity and statistics are the topic of Sect.~\ref{sect_structure}, and I conclude in Sect.~\ref{sect_summary} with an overview over future developments, caveats and a summary of the implications of cosmology for fundamental physics. The article summarises my contribution to the workshop "Why trust a theory?" taking place from the 07.Dec.2015 to the 09.Dec.2015 at the Ludwig-Maximilians-Universit{\"a}t M{\"u}nchen.

\section{gravity and general relativity}\label{sect_gravity}
Adopting the principle of extremised proper time as a generalisation of the principle of least action from classical mechanics to the relativistic motion in gravitational fields leads to the construction of metric theories of gravity \citep{einstein_feldgleichungen_1915, einstein_grundlage_1916}: The equivalence principle stipulates that all objects fall at the same rate in gravitational fields irrespective of their mass, which suggests that the proper time is affected through the metric as the expression of the gravitational field. In the absence of gravitational fields (or in an infinitesimally small freely falling and non-rotating laboratory without the possibility of observing distant charges) the metric assumes the Minkowskian shape, while the first derivative of the metric vanishes, reducing the covariant derivative to a partial one, such the dynamics of all systems is determined locally by the laws of special relativity. The vanishing first derivatives of the metric is the reason, from a relativistic point of view, why pendulum clocks stop ticking in free fall.

The clearest physical setup in which one can experience gravitational fields is that of geodesic deviation, i.e. the relative motion of two test particles which fall in a gravitational field: The second derivative of the distance between two test particles with respect to proper time is proportional to the Riemann-curvature. In particular, if the curvature is zero, the test particles increase or decrease their distance at most linearly, but if the curvature is non-vanishing, one can see more complicated forms of relative motion, for instance the acceleration or deceleration of distant galaxies in cosmology or the oscillatory motion of two objects in the field of a gravitational wave. 

The metric as a generalisation to the gravitational potential is sourced by the energy-momentum density in a field equation, and the gravitational field has a relativistic dynamic on its own which is described by the (contracted) Bianchi-identity. It is very impressive that the shape of the field equation is completely determined by fundamental principles and that "God did not have any choice in the creation of the world" as Einstein put it -- although it was only rigorously proved by Cartan \citep{cartan_sur_1922}, Vermeil \citep{vermeil_notiz_1917} and Lovelock \citep{lovelock_uniqueness_1969} that the general relativity is the only possible metric theory of gravity based on Riemannian differential geometry in four dimensions which is based on a single dynamical field, which is locally energy-momentum conserving, with a local field equation, and which is based on derivatives up to second order \citep{lovelock_einstein_1971, navarro_lovelocks_2011}.

It is worth to discuss these five requirements in detail because they reflect fundamental principles of physical laws, which makes gravity so interesting: Because there is no alternative theory of gravity under these assumptions, any deviation from general relativity will directly call these fundamental principles into question. Cosmology plays an important role in this context because it is the most simple gravitational system which, in contrast to the Schwarzschild-solution or gravitational waves in vacuum, involves the field equation where the local Ricci-curvature is directly coupled to the cosmological fluids.

A relation between the second derivatives of the metric and and the source of the gravitational field is necessary because of a number of reasons. Apart from analogies to other relativistic field theories like electrodynamics, which feature identical constructions, second-order field equations are a very natural way to incorporate parity and time-reversal symmetry in a field theory, and in fact general relativity is perfectly $\mathcal{PT}$-symmetric, in particular in the weak field limit, where this symmetry is not already provided by general covariance. Secondly, the definition of curvature is based (in the simplest constructions of parallel transport) on the non-commutativity of the second covariant derivatives, which ensures that the Bianchi-identity applies to the description of the propagation dynamics of the gravitational field and enforces local energy momentum conservation. Lastly, the inclusion of higher-order derivatives would lead to the Ostrogradsky-instability \citep{woodard_theorem_2015} because systems are not necessarily bounded energetically from below, although there might be constructions which lead to stable theories despite the inclusion of higher derivatives. And from a fundamental point of view it is natural to base a gravitational theory on second derivatives of the metric because of their invariant nature, as tidal fields cannot be set to zero by transforming into freely falling frames.

Arguing from fundamental physics adds a very interesting thought: The dynamics of fields is modelled on the dynamics of (relativistic) particles, for instance one uses concepts like the action as a generalisation of the principle of extremised proper time in analogy for the causal transition from one field configuration to another. Lagrange-functions for point particles are convex due to causality and yield again convex Hamilton-functions which are bounded from below, resulting in stability. The construction of a convex Lagrange-function involving squares of first derivatives is naturally suggested by special relativity. This construction principle applies equally to the gravitational field, where the formulation by Einstein and Palatini involving squares of Christoffel-symbols is  natural from this point of view \citep{palatini_deduzione_1919}, and yields in addition to the field equation the metric connection. Assuming this relationship as an axiom allows the Einstein-Hilbert-action as the second formulation in which general covariance and the locality of the field equation are apparent.

The four dimensions of spacetime are peculiar: While classical, Newtonian gravity can be formulated in any number of dimensions, general relativity requires at least four dimensions, which can be seen from a number of arguments, the most elegant perhaps being the fact that the Weyl-tensor, which contains the non-local contributions to the Riemann-curvature and which describes the propagation of the gravitational field away from its sources, is only nonzero in four or more dimensions. If one assumes a larger number of dimensions one needs immediately to provide an explanation why there are differences between the basic four dimensions and the additional ones, in particular in invoking the equivalence principle which locally provides a reduction to flat Minkowski-space which has four dimensions. Conversely, there are empirical requirements which limit the number of dimensions: For instance, there are no stable Kepler-orbits in more than four dimensions and the Huygens-principle is only exactly valid in spaces with an odd number of spatial dimensions.

Locality of the field equation is realised in general relativity because only the Ricci-curvature is related to its source, the energy-momentum-tensor, while the Weyl-tensor is unconstrained. The fact, that the Riemann-curvature is only partially determined by the field equation and that the Bianchi-identity as a second ingredient for the field dynamics is the construction that allows the field to propagate away from its sources, realising Newton's definition of a field as "an action at a distance". 

The origin of energy-momentum conservation are symmetries of the Lagrange-density which determine the internal dynamics of the gravitating substance. If the dynamics of the substance is universal and does not depend on the coordinates, the Lagrange-density defines together with the metric a conserved energy-momentum-tensor: In this sense, general relativity is the gravitational theory for locally energy-momentum-conserving systems in the same way as electrodynamics is a field theory for charge-conserving systems, and one notices a remarkable consistency between the dynamics of the fields and the dynamics of the field-generating charges. This is realised in both gravity and electrodynamics in the antisymmetry of the curvature form.

Finally, general relativity uses the metric as the only dynamical degree of freedom of the gravitational field. While a larger number of gravitational degrees of freedom are conceivable \citep{de_rham_massive_2014} and form the basis of the Horndeski-class of gravitational theories \citep{horndeski_second-order_1974}, one must include an explanation why the gravitational interaction of all particles is exclusively determined by the metric and not by any other of the gravitational degrees of freedom, because of the unique interpretation of the metric in the principle of extremised proper time. Apart from the uniqueness of the metric, this point seems the easiest assumption to overcome in comparison to the four others. Personally, I would be interested to know whether the introduction of new gravitational degrees of freedom has implications in relation to Mach's principle, which was the primary motivation for Brans and Dicke to introduce a scalar degree of freedom in their gravitational theory \citep{brans_machs_1961}.

It is remarkable that general relativity is the only consistent expression of a gravitational theory based on these five principles, and that they equally apply to other field theories, for instance to electrodynamics, which is likewise unique as a $\mathcal{PT}$-symmetric, charge-conserving, local field theory with a single, vectorial field in four dimensions, for exactly the same reasons, with the only difference being the replacement of general coordinate covariance by Lorentz covariance. It is difficult to overstate the implications of this uniqueness, because any deviation from the predicted behaviour of the theories questions the fundamental assumptions for their construction. And in addition, the cosmological constant $\Lambda$ is, under these assumption, a necessary feature of the field equation, although the metric exhibits a different Weyl-scaling than the Einstein-tensor - from this point of view the cosmic coincidence is more remarkable, because the transition between the two scaling-regimes takes place in the current epoch.

The Universe is a unique testbed for strong, geometric gravity with the involvement of the field equation in a very symmetric setup with pure Ricci-curvature, and a system which is large enough to display order-unity effects in relation to $\Lambda$. Currently, most of the research focuses on additional dynamical degrees of freedom of gravity as well as a possible non-locality of the interaction with the gravitational field source, where the influence of gravity on cosmic structure formation constitutes the physical system \citep{clifton_modified_2012-1, koyama_cosmological_2016, joyce_dark_2016}.

\section{cosmology as an application of gravity}\label{sect_cosmology}
Cosmology provides a natural example of strong-field, geometric gravity in a very symmetric setup. Under the Copernican principle, which assumes spatial homogeneity and isotropy, the Friedmann-Lema{\^i}tre-Robertson-Walker-class of cosmologies exhibit the same degree of symmetry as the exterior Schwarzschild-solution or the gravitational waves - these three cases are discussed in every textbook, because they allow exact solutions to the gravitational field equations due to their high degree of symmetry. In fact, if the degree of symmetry were any higher, there could not be curvature, and consequently no effects of gravity.

There is, however, one very important difference between these three solutions: Whereas the Schwarzschild-solution and gravitational waves are vacuum solutions and exhibit curvature in the Weyl-part of the Riemann-tensor, the curvature in FLRW-cosmologies is contained in the local Ricci-part. As an immediate consequence one needs the full field equation for linking the geometry and the expansion dynamics of the FLRW-cosmologies to the properties of cosmological fluids, which is in contrast to the Schwarzschild-solution or the gravitational waves, where the Bianchi-identity is sufficient in its property of describing the dynamics of curvature. To put this in a very poignant way, cosmology investigates gravity in matter, and because the curvature is in the Ricci-part of the Riemann-tensor due to the symmetry assumptions of the metric, one does not have to deal with propagation effects of gravity unlike in the case of the vacuum solutions.

Equating the Einstein-tensor with the energy-momentum-tensor of a homogeneous, ideal relativistic fluid leads to the Friedmann-equations, which describe acceleration or deceleration of the cosmic expansion in its dependence on the density and the pressure-density relation (i.e. the equation of state) of the cosmic fluids. Under the symmetry assumptions of the Copernican principle only the density and the equation of state can characterise a cosmic fluid as a source of the field equation. As a consequence of the field equation, which relates geometry to energy-momentum density, it is possible to assign these two properties as well to geometric quantities like spatial curvature or to the cosmological constant, although they do not correspond to actual physical substances: For instance, the density parameter associated with curvature can in principle be negative, and the density associated with $\Lambda$ does not decrease with increasing scale-factor. In these cases the assigned density parameter and the equation of state only characterise the influence on the expansion dynamics in contrast to an actual substance as a classical source of the gravitational field, where the equation of state is related to the relativistic dispersion relation of the particles that make up the fluid.

The cosmological constant is a natural term in the gravitational field equation under the conditions discussed in the last section, but it only influences the expansion dynamics of the Universe on scales in excess of $10^{25}$ meters, i.e. on today's Hubble scale $c/H_0$, due to its numerical value. The numerical similarity of the density parameters $\Omega_m$ and $\Omega_\Lambda$ constitutes the coincidence problem because spatial flatness only requires their sum to be close to one, $\Omega_m+\Omega_\Lambda=1$, but there is no argument on the basis of which one can make statements on the individual values, in particular because they show a very different evolution with time.

Spatial curvature, i.e. the curvature of spatial hypersurfaces at fixed time, is a degree of freedom which is allowed by the Copernican principle, but which is not realised in Nature at the current epoch. While there is no symmetry which fixes curvature to be zero, cosmological observations constrain the associated density parameter to be at most a percent, or put a lower limit of the curvature scale to be at least $\simeq 10c/H_0$ \citep{planck_collaboration_planck_2015-5}. The problem of small spatial curvature is exacerbated by the fact that curvature grows if the cosmic expansion is decelerated, and it was in fact decelerating for most of the cosmic history under the dominating gravitational effect of radiation at early and of matter at later times. A widely accepted mechanism which explains the smallness of spatial curvature is cosmic inflation, where spatial curvature is made to decrease by assuming an early phase of accelerated expansion.

Likewise, it is possible to assign density and pressure to the energy-momentum tensor of fields with self-interactions which form the basis of quintessence models of dark energy \citep{mortonson_dark_2013-1}: In those models one obtains a conserved energy-momentum density from the coordinate independence of the Lagrange-density. Identification of terms leads to relations between the fluid properties density and pressure to the kinetic and potential terms of the field, and constrains the equation of state to a physically sensible range which in particular leads to finite ages of the cosmological models. If the dark energy density is chosen to be comparable to that of matter and if the equation of state is negative enough one observes as the most important effect accelerated expansion at late times \citep{huterer_probing_2001,frieman_dark_2008}.

Geodesic deviation is again a very illustrative example of how gravity affects the motion of galaxies in cosmology: Each galaxy is considered to be a freely falling test particle whose relative acceleration depends on the spacetime curvature, which in turn is linked through the field equation to the density of all cosmological fluids and their equations of state. On distances in excess of $c/H_0$ one can see that galaxies increase their distance relative to the Milky Way in an accelerated way under the action of the cosmological constant, the presence of dark energy or through a modification of the gravitational field equation. Naturally, light from these distant galaxies reaches us on our past light cone as photons move along null-geodesics, and therefore the measurement in this case consists in relating the distance measured on the past light cone to the redshift of the galaxies. It is essential that measurements in cosmology involve the relative motion of macroscopically separated objects: Any local, freely-falling and non-rotating experiment will experience the metric as being locally Minkowskian, and would conclude from the absence of curvature that there are no gravitational field generated.

One issue, which defies a definitive formal solution and which is still a matter of current debate is the problem of backreaction \citep{buchert_is_2015, bolejko_inhomogeneous_2016, adamek_general_2013, green_examples_2013, wiltshire_cosmic_2007}: Due to the nonlinearity of the gravitational field equations it matters how the gravitational field on large scales, where statistical homogeneity and isotropy should apply, arises from the structured matter distribution on small scales, in other words, how the Weyl-curvature due to structures on small scales transitions to the Ricci-curvature of the homogeneous FLRW-universe. There are analytical arguments as well as simulations of the spacetime geometry, but a definitive answer is still not reached.

\section{$\Lambda$CDM as the standard model}\label{sect_lambdacdm}
Modern cosmology has established $\Lambda$CDM as its standard model \citep{silk_challenges_2016-1}: In this model, the expansion dynamics of the Universe is described by general relativity with the assumption of the Copernican principle, leading to the FLRW-class of homogeneous and isotropic cosmologies with very small spatial curvature and with radiation, (dark) matter and the cosmological constant $\Lambda$ as dominating, ideal fluids at early, intermediate and late times, respectively. The confirmation of spatial flatness and accelerated expansion was achieved by combining supernova data with observations of the cosmic microwave background, which determine to leading order the sum and difference of the matter density parameter and the density parameter associated with $\Lambda$, while there are indications for isotropy from the Elhlers-Geren-Sachs-theorem applied to the cosmic microwave background, even in the case of small perturbations \citep{ehlers_isotropic_1968, stoeger_proving_1995, clarkson_inhomogeneity_2010, planck_collaboration_planck_2016}

While the density parameters are well determined by observations and even the equation of state parameters have been constrained to values suggesting that the cosmological constant is the driving term behind accelerated expansion, the fundamental symmetries of the FLRW-cosmologies are only partially tested: While tests of isotropy provide support for the Copernican principle, homogeneity is difficult to test and there are weak indications that it is in fact realised in Nature, for instance from the magnitude of the Sunyaev-Zel'dovich effect in Lema{\^i}tre-Tolman-Bondi-models. In a certain sense, deviations from homogeneity does not lead to a less complex cosmological model: After all, we as observers need to be positioned close to the centre of a large, horizon-sized void.

The bulk of gravitating matter is required to be non-interacting with photons (from the amplitude of structures seen in the CMB and their growth rate in comparison to the amplitude of large structures today), to have a very small cross-section for elastic collisions (leading to be pressureless and inviscid "fluid" which is responsible for the typical core structure of dark matter haloes), to have very little thermal motion, i.e. to be dynamically cold (from the power-law behaviour of the halo mass function at small masses and lately from the abundance of substructure in dark matter haloes). 
\citep{}\citep{}
Fluctuations in the distribution of matter and in the velocity field at early times are thought to be generated by cosmic inflation \citep{martin_encyclopaedia_2013-1, tsujikawa_introductory_2003, baumann_cosmological_2009}, which at the same time remedies the flatness problems discussed earlier, and provides in addition an explanation of the horizon-problem, i.e. the uniformity of the CMB on large scales. Cosmic inflation generates fluctuations whose fluctuation spectrum follows a near scale-free law known as the Harrison-Zel'dovich spectrum, and generated mostly adiabatic perturbations, i.e. identical fractional perturbation in every cosmological fluid, with near-Gaussian statistical properties, which has been tested most importantly with observations of the cosmic microwave background.

In summary, $\Lambda$CDM as the standard model for cosmology is a model with low complexity but makes extreme assumptions at the same time, most importantly the Copernican principle and the properties of dark matter, and it features with cosmic inflation a construction in order to explain the origin of fluctuations as well as the absence of spatial curvature. 

Adopting the view that general relativity including the cosmological constant is the only possible metric theory of gravity under certain concepts, allows to estimate scales by which the Universe should be correctly characterised. Constructing a length scale $l_H=1/\sqrt{\Lambda}$, a time scale $t_H=1/(c\sqrt{\Lambda})$ and a density scale $\rho_H=c^3/(G\sqrt{\Lambda})$ from the speed of light $c$, the gravitational constant $G$ and the cosmological constant $\Lambda$ leads to values which to a very good approximation correspond to the Hubble length $c/H_0$, the Hubble time $1/H_0$ and the critical density $\rho_\mathrm{crit} = 3H_0^2/(8\pi G)$, if the curvature scale becomes infinitely large, or equivalently, the associated density parameter $\Omega_K$ approaches zero.

Comparing these scales to the Planck-system of units with $l_P=\sqrt{G\hbar/c^3}$, $t_P=\sqrt{G\hbar/c^5}$ and $\rho_P=c^5/(G^2\hbar)$ constructed from the speed of light $c$, the gravitational constant $G$ and the Planck constant $\hbar$, shows large discrepancies, most notably the estimate $\rho_P = 10^{120}\rho_H$, which is usually stated as a motivation for dark energy models: If one associates the effects of the cosmological constant to the expectation value of the field theoretical vacuum, one obtains values which are much too large in comparison to $\rho_H$, but the problem seems to be constructed because it would apply to matter in exactly the same way: The matter density $\Omega_m\rho_\mathrm{crit}$ is close to $\rho_H$ instead of $\rho_P$. Personally, I am lost at the fact that with the inclusion of the cosmological constant $\Lambda$ as a natural constant to general relativity (required by Lovelock, Cartan and Vermeil) the definition of a natural system of units is not unique, and that furthermore the early universe is described by one choice of constants (namely $c$, $G$ and $\hbar$) and the late, current universe by the other possible choice ($c$, $G$ and $\Lambda$). Adding the Boltzmann-constant $k_B$ to this argument one can show that the number $10^{120}$ is the smallest non-trivial number that can be constructed from all fundamental constants of Nature.

In a certain sense, $\Lambda$CDM suffers from being a successful standard model, because it is difficult to find evidence for its failing, which is of course very motivating and fascinating. But in this respect it is necessary to report all statistical tests irrespective of whether they provide evidence for or against $\Lambda$CDM, in the light of all available data. And perhaps one would make clearer whether a certain model construction is part of the theoretical discussion or thought as a viable alternative. From what was discussed in the previous section, it might as well be that progress comes from conceptual work on gravity, in analogy to the construction of general relativity in the first place over a century ago.

\section{statistics in cosmology}\label{sect_statistics}
As an astronomical discipline, observational cosmology restricts the observer to a purely passive role, because it is impossible to have an active influence on the dynamics of the cosmological expansion or the formation of cosmic structures. Because of this impossibility of carrying out experiments one is forced to require that only questions can be answered for which experiments have been carried out by Nature. Concerning the expansion dynamics this does not seem like a large restriction because we can observe the behaviour of FLRW-cosmologies over a wide range of values for the density parameters and for dominating fluids with different equation of state parameters. Certain models, for instance phantom dark energy models where a very negative equation of state parameter leads to a infinite scale factor after a finite time, do not seem to be realised in Nature even though the gravitational dynamics of such a model would be very interesting to investigate. Similarly, the investigation of small values of curvature is difficult because the finite horizon size does not permit to access large enough scales.

The passive role of cosmological observers is a central issue in the observations of cosmic structures \citep{coles_statistical_2003}. There, one assumes that the random process of which the actual distribution of matter in our Universe is one realisation, is ergodic, meaning that we can determine ensemble-averaged quantities which characterise the random process, by performing spatial averages over data taken from inside our horizon. Ergodicity can not be experimentally verified, but there are strong theoretical indications that it is a good assumption. On theoretical grounds, a theorem by R.J. Adler stipulates that ergodicity is ensured in Gaussian random fields with continuous spectra. Ergodicity is an important assumption when investigating objects such as clusters of galaxies, because only sufficiently large volumes yield statistically unbiased samples because of a large enough set of initial conditions from which the objects have formed. On smaller scales, there are theoretical investigations that relativistic effects in surveys related to the light propagation cause a violation of ergodicity on a small level.

While the statistical properties of the cosmic large-scale structure are in fact very close to Gaussian in the early universe or today on large scales, nonlinear structure formation generates large deviations from Gaussianity, which are usually quantified in terms of $n$-point correlation functions or polyspectra of order $n$ in the case of homogeneous random processes. It is unknown if there is an efficient description of non-Gaussianity introduced by nonlinear structure formation. Up to now one can predict all higher-order correlators in perturbation theory where they are sourced due to the nonlinearities in the continuity- and Euler-equations. Although many alternatives have been explored, for instance extreme value statistics or Minkowski functionals, which might in principle provide an efficient quantification of non-Gaussianity, their relationship to the fundamental dynamical equations is less clear than in the case of correlation functions.

All estimates of statistical quantities suffer from cosmic variance, i.e. they are limited in their precision by the law of large numbers: Any moment of a random variable can only be estimated with a finite variance which scales inversely proportional to the number of samples. In cosmology, this limits the detectability of correlation functions or spectra on large scales in a statistical way, even if the estimate itself is unbiased: In the measurement of angular spectra of the galaxy density, of the weak lensing signal or of the temperature of the cosmic microwave background there are only a finite number of modes available at a given angular scale, such that spectra can only be determined at an uncertainty inversely proportional to the angular scale. This limits cosmology in a fundamental way in particular on the largest scales.

The cosmic large-scale structure is to a very good approximation described by classical fluid mechanics on an expanding background on scales smaller than the Hubble-scale $c/H_0$ and in the linear regime of structure formation. The dynamical variables are the density and velocity field, if a fluid mechanical approach is adopted - which is conceptually not correct because of the collisionlessness of dark matter, which is required to have a very small, possibly vanishing cross-section for elastic collisions. The phase-space information contained in the density and velocity field are accessible at different degrees of precision into the directions parallel and perpendicular to the line of sight. A grand result which should be mentioned in this context is the detection of baryon acoustic oscillations in the distribution of galaxies, which links the large-scale structure at a very early point in cosmic history to the present time and shows that the statistics of the cosmic matter distribution is conserved in linear structure formation.

Counting the cosmological parameters associated with the FLRW-dynamics (the matter density $\Omega_m$, the dark energy density $\Omega_w$, which is replaced by $\Omega_\Lambda$ in the case of the equation of state being $w=-1$, the curvature $\Omega_K$, which assumes very small values due to cosmic inflation, and the Hubble-parameter $h$ itself) and those describing structures, i.e. fluctuations in the distribution of matter (the fluctuation amplitude $\sigma_8$ and the slope $n_s$ of the spectrum) results in six parameters, seven if a time evolution of the dark energy equation of state is admitted. The spectral slope $n_s$ and the amplitude $f_\mathrm{NL}$ of non-Gaussianities have to result consistently from the inflationary slow-roll parameters. As cosmological probes usually combine information from the homogeneous expansion dynamics with cosmic structures, measurements depend on a large set of at least seven physical parameters \citep{np_standard_2001}.

\section{inference in cosmology}\label{sect_inference}
The process of inference, i.e. the estimation of model parameters (or at least of bounds on model parameters) from cosmological data differs substantially from other parts of physics: It is close to impossible to investigate certain aspects of the cosmological model independently from others. Contrarily, in elementary particle physics direct detection experiments measure e.g. mass, charge or spin of particles and their couplings independently from other aspects of the standard model of particle physics, and only in indirect detection experiments for probing physics above the provided energy one needs to deal with simultaneous contributions from many physical processes and particles.

Any large-scale structure observable and its statistical properties depend in the most straightforward case on the density parameters of all cosmic fluids and their respective equations of state as well as on parameters which characterise the distribution of matter. Cosmological observations combine statistical properties of the fluctuations with their growth rates and the relation between observable redshift and distance. The investigation of models as complex as those in cosmology require very strong signals, for instance the spectrum of CMB temperature fluctuations or the weak lensing signal of galaxies constitute signals with a statistical significance of the order of $10^3\sigma$. There is a natural progression in cosmology from the observation of the homogeneous and isotropic expansion dynamics on large scales (e.g. with supernova-measurements) to the statistics of linear fluctuations (in the cosmic microwave background) and finally to the evolved, nonlinear cosmic structure on small scales and at late times.

In addition to that, one needs to deal in most cases (with the possible exception of weak lensing) with a model which relates the observable to the underlying dark matter objects, for instance biasing relation for the density of galaxies in relation to the dark matter densities or mass-temperature-relations for the luminosity of clusters or their Sunyaev-Zel'dovich or $X$-ray signal. If the set of cosmological parameters along with effective models for the observables is to be constrained by data, large degeneracies are natural and many algorithmic sampling techniques have been developed to treat likelihoods in high-dimensional parameter spaces. Sampling algorithms based on Monte-Carlo Markov-chains are optimised to deal with strong degeneracies and can sample at high efficiency from strongly non-Gaussian likelihoods with an implicit assumption about uni-modality of the likelihood.

A further complication is hidden in the hierarchy in sensitivity towards cosmological parameters which is inherent to cosmology: Certain parameters are more difficult to measure than others, which means that a measurement needs to provide exquisite precision on easy parameters in order to provide interesting statements about parameters which are difficult to measure. Examples might include weak lensing, which is strongly dependent on the matter density $\Omega_m$ and the fluctuation amplitude $\sigma_8$, and weakly dependent on the dark energy equation of state parameters $w_0$ or its time evolution $w_a$, such that one needs to achieve relative accuracies of $10^{-4}$ on $\Omega_m$ and $\sigma_8$ in order to answer questions in relation to dark energy, for instance the equation of state-parameters $w_0$ and $w_a$. A similar situation is found in the interpretation of data of the cosmic microwave background, where one central parameter is the optical depth $\tau$, which needs to be constrained for fundamental parameters to be accessible.

It is breathtaking that the future generation of surveys will constrain parameters of a $\Lambda$CDM-type cosmology or of a basis dark energy-cosmology to a level of $10^{-4}$, in particular because Newton's gravitational constant is known at similar levels of precision from laboratory experiments. At this level, the difficulty will lie in a detailed understanding of the relationship between the observable quantity and the fundamental cosmological model. On small scales, this involves an understanding of astrophysical processes governing the properties of observed objects and their relation to the ambient large-scale structure as well as nonlinear corrections to observables and a treatment of their covariances. For instance, gravitational lensing, recognised as a very clean probe with a straightforward functioning principle which is entirely determined by relativity, requires a precise prediction of second-order corrections to lensing, a modelling of intrinsic alignments of galaxies and an accurate description of the statistical distribution of estimates of the spectrum. Quite generally one can expect that cosmology in the next generation of experiments will not be limited by statistical but rather by systematical errors.

If one is in the situation of deciding between two viable theoretical models one would like to set up a decisive experiment which is designed in a way to favour one of the models strongly over the other \citep{trotta_bayesian_2017}. It is difficult to pursue this idea in cosmology, because one can at most $(i)$ use an observational technique in which the difference between two models ist most pronounced, or $(ii)$ optimise an experiment in such a way that the two models in question show a large difference in their generated signal, or $(iii)$ use degeneracy-breaking combinations of probes.

These optimisations can only show quantitative differences due to the degeneracies involved, such that the likelihoods of models differ not very strongly in parameter space and typically show tensions at the level below $3\sigma$, making a decision between models difficult. Because of this many methods for model comparison have been developed, which quantify the trade-off between models of different complexity in their ability to fit data and in their simplicity. In this sense, methods of Bayesian model comparison are a quantification of Occam's razor, preferring simpler models over more complex ones if both models fit data similarly well, but decide in favour of more complex models only if the data requires a higher complexity. Specifically, Bayesian evidence measures the probability that a model is able to explain the data integrated over the model's parameter space weighted with distribution which quantifies the knowledge on the parameters prior to the experiment. Two competing models are compared by the ratio of their Bayesian evidences, which in turn can be quantified by e.g. the Jeffrey's scale for a quantitative interpretation of the Bayesian evidence ratio, although the scale for the degree of confidence in a model is, like in all applications of statistical testing, arbitrary. But on the other hand it might be worth realising that complexity of models is a human concept and we commonly assume that the laws of Nature are conceptually simple (but technically complicated at times!).

Although it might be a weird thought at first, it is worth noticing that the future generation of experiments will survey such large volumes that they harvest a significant part of the cosmologically available information, and that there is a natural limit in the complexity of cosmological models which can at best be investigated. Therefore, it is imperative to develop new techniques and new probes to open up new windows, for instance to structures in the high-redshift universe in 21cm-observations, or the observation of recombination lines from the epoch of atom formation. Alternatively, there is the fascinating possibility of real-time cosmology, where one directly observes changes in the recession velocity of distant quasars in cosmologies with an effective equation of state unequal to $w=-1/3$, or at least in principle, small changes in the fluctuation pattern of the cosmic microwave background on the time scale of 100 years. An alternative route to accessing scales much smaller than typical galaxy separations or the scales responsible for the substructure in haloes is the search for primordial black holes, whose abundance is determined by the inflationary model.

\section{structure formation}\label{sect_structure}
One of the goals of cosmology is a theory for the formation and evolution of cosmic structures, with the ultimate aim of providing a quantitative theory of galaxy formation, including a physical picture of galaxy properties, their evolution and their relation to the cosmic large-scale structure \citep{frenk_dark_2012}. Due to nonlinearities in the underlying system of equations structure formation is a difficult and not yet fully understood problem even for dark matter alone, which obeys comparatively straightforward physical laws: For predicting the scale-dependence and time-evolution of statistical quantifiers one carries numerical simulations, where algorithmic progress has allowed to run simulations with $8192^3$ particles \citep{alimi_deus_2012}, or uses perturbative methods \citep{bernardeau_large-scale_2002}, where recent developments have made fully analytic treatments possible \citep{bartelmann_microscopic_2014}. The inclusion of baryonic matter into structure formation is realistically restricted to numerical simulations, where, depending on the physical processes considered, one deals with a multi-scale and multi-physics problem whose results are still influenced by the choice of algorithms for solving the coupled system of differential equations, the discretisation scheme and the details of the implementation of physics on small scales \citep{genel_introducing_2014, springel_e_2010}.

In the process of increasing the data volume it is obvious that much of the signal collected by future large-scale structure surveys is generated by nonlinearly evolving structures. For instance, the weak lensing signal measured by the future Euclid-mission will amount to $>10^3\sigma$ of statistical significance, of which about three quarters are generated by nonlinear structures. Clearly, this requires a detailed understanding of phenomena related to nonlinear structure formation, in particular the inter-dependence of statistical quantifiers. While structure formation in the linear regime, i.e. for small perturbations, is easily shown to preserve all statistical properties of the initial conditions, which is the reason why it is possible to investigate inflationary theories with observations of the cosmic microwave background or of the cosmic structure on large scales, nonlinear structure formation generates an entire hierarchy of higher-order cumulants in a coupled system. Perturbative approaches are in principle possible, but technically very difficult, and it is only lately that progresses in effective field theory and statistical field theory provide a theory of statistical evolution of the large-scale structure agreeing with numerical simulations on small scales.

The internal properties of haloes, in particular their rotation curves and their virial equilibria, provide support for dark matter to be gravitationally dominating, and from the core structure of haloes and from the abundance of substructure one places constraints on the collisionlessness and its thermal motion, respectively \citep{colafrancesco_dark_2010-1}. Additionally, structure growth in the early universe clearly requires dark matter to be non-baryonic. Haloes have been found in numerical simulations to have a certain profile shape, which is universal, scales weakly with halo mass and naturally accommodates the constant rotation velocity at large distances from the halo centre. In addition, observations from weak gravitational lensing show that the profile shapes of haloes correspond well to those found in numerical simulations although there is not yet a complete physical understanding of their stability and of the origin of their universal shape from first principles. Although there are very strong arguments for the existence of dark matter from cosmology and from astrophysics, the relation to elementary particle physics is still unclear: Many particle candidates are being discussed, from very light particles like axions with a mass of $\simeq 10^{-22}~\mathrm{eV}$ to WIMPs close to the supersymmetric scale at a few $\mathrm{TeV}$ \citep{bergstrom_dark_2009, fornengo_dark_2016, queiroz_dark_2016}.

Cosmic inflation is a construction which provides a natural framework for the generation of cosmic structures, which in the course of cosmic history grow by gravitational instability. Despite all the successes of inflation it must be emphasised that it is a construction involving non-gravitational physics to explain the symmetries of the FLRW-cosmology as an entirely gravitational system, and that its predictions follow from the construction of the particular inflationary model and not generically from fundamental physics. Apart from the scale-dependence of the fluctuation amplitude cosmic inflation generates small non-Gaussianities which are dwarfed by nonlinear evolution on small scales and are measurable on very large, quasi-linearly evolving scales at high redshift. These small amounts of non-Gaussianity reflect derivatives of the inflationary potential and should be consistent with the slow-roll parameters which determine the scale dependence of the inflationary fluctuations \citep{planck_collaboration_planck_2016-1}. It seems to be difficult, but not fundamentally excluded, to construct models with large non-Gaussianity. An independent measurement of the two slow-roll parameters and the corresponding non-Gaussianity parameters is yet to be carried out.

It is difficult to overstate the importance of computer simulations of cosmic structure formation for modern cosmology: Due to the nonlinearity of the structure formation equations, with many different active physical processes and the difficulty of separating physical scales. The sophistication of numerical simulations has led to a framework for dark matter driven structure formation from small to large scales, the thermal evolution of the baryonic component, leading to galaxy formation and star formation on small scales and is able to predict many properties of galaxies with the dependence on time, statistics, scaling behaviour and morphology. The development of analytical methods was not quite commensurate with this evolution, because analytical methods are mostly constrained to the dark matter component and focus on predicting the evolution of non-Gaussianities in a perturbative way, and a detailed theoretical understanding of many numerical results is still missing, for instance concerning the stability of dark matter haloes. The worry that in cosmology one is accepting to be strongly dependent on numerical simulations is justified, both for high-precision predictions of linear perturbations, in particular with more involved particle models, as well as predictions for nonlinear structures. But this is not very different compared to elementary particle physics, where the search for particles outside the standard model or at high energies involve field-theoretical computations with similar complexity. Finally, there is always the question of interpreting simulations, isolating individual structure formation processes and quantifying degeneracies in model parameter choices.

\section{Conclusion}\label{sect_summary}
Modern cosmology relies on three building blocks: general relativity as a theory of gravity, (fluid) mechanics as a theory of motions in the distribution of matter and statistics for their description. It offers views on gravity as one of the fundamental forces, both for the dynamics of the Universe on large-scales as well as the driving force for the formation of cosmic structures. The universe is a system for testing the gravitational interaction on large scales in a fully-relativistic, yet reasonably symmetric system, with new gravitational effects in relation to the cosmological constant $\Lambda$. The dark matter distribution in the universe does not form a continuum, and for that reason the usage of fluid mechanics is conceptually inadequate, in contrast to the baryonic component.

General relativity as a theory of gravity follows from a number of fundamental physical principles and is unique as a metric theory under certain assumptions. It includes naturally the cosmological constant $\Lambda$ as a term in the field equation and any measured deviation of the behaviour of gravity would question these fundamental principles. A consequence of assuming general relativity as the theory of gravity is the need for the existence of dark matter, which is substantiated by its role in structure formation and by its influence on the internal properties of dark matter structure, but whose relationship to particle physics is still unclear.

There are many interesting directions in which cosmology is currently evolving: Large-scale surveys cover significant fractions of the observable cosmic volume, and it is imperative to develop new probes, for instance observations of the large-scale structure with the hyperfine transition of neutral hydrogen, which makes the high-redshift universe accessible \citep{bull_measuring_2015, kitching_euclid_2015, maartens_cosmology_2015} and shows structures in a close to linear stage of evolution . Another frontier are $B$-modes of the polarisation of the cosmic microwave background, which are directly related to the energy-scale of inflation. As a completely different approach, changes in cosmological signals with time can be investigated. Many current cosmological probes are limited by astrophysical processes on small scales which are difficult to model and require simulations with a correct implementation of all relevant physical processes. If these limitations can not be controlled, cosmology will be limited by systematical rather than statistical errors. On the other hand, Nature provides very clean standard signals in the form of the gravitational wave signals of coalescing black hole binaries, which have the potential of surpassing supernovae in precision. Finally, one could dream of independent measurements of the slow-roll parameters and of finding consistency between the predictions of inflation in the shape of the spectrum, the amounts of non-Gaussianity and the level of primordial gravitational waves.

At the stage of interpreting cosmological data it should be kept in mind that according to the principle of Bayesian evidence simpler models should be preferred over more complex ones, and one should accept more complex models only if they provide substantially better fits to data. While conceptually clear, it might be very difficult to apply this as one has to deal in the next generation of experiments with very large data volumes and models which not only include physical parameters but need to be sufficiently flexible to cope for astrophysical peculiarities of the probes employed.

On the theoretical side it would be important to have an effective (and computationally efficient) theory of nonlinear structure formation with a prediction and an efficient description of non-Gaussianity, and possibly incorporating relativity into the prediction of structure growth. These analytical methods compete with very evolved and sophisticated numerical simulations for structure formation, which move in their aim away from fundamental cosmology to a quantitative model for galaxy formation and evolution.

I hope that this article could summarise some of the fundamental questions currently being investigated in cosmology and illustrate their relationship to the principles how the laws of Nature are formulated. Many aspects of cosmology rely on arguments from simplicity and can never be rigorously tested. Keeping in mind that physical cosmology is only 99 years old, counting from the discovery of redshifts of galaxies by Slipher soon after the formulation of general relativity, it has come very far in linking physical processes at the Hubble-scale to those close to the Planck-scale, and I am confident that many surprises wait for their discovery.

\section*{Acknowledgements}
I would like to thank Matthias Bartelmann, Richard Dawid, Yves Gaspar, Frank K{\"o}nnig, Karl-Heinz Lotze, Sven Meyer, Robert Reischke and Norman Sieroka for very constructive and valuable comments on the draft.

\bibliographystyle{mnras}
\bibliography{references}

\bsp
\label{lastpage}
\end{document}